\documentclass[12pt]{article}
\usepackage{amsmath}
\usepackage{amssymb}
\usepackage{amsthm}
\makeatletter
\@addtoreset{equation}{section}
\makeatother

\newcommand{\pa}{\partial}
\newcommand{\bm}[1]{\textnormal{\mathversion{bold}$#1$}}

\newcommand{\defeq}{\stackrel{\mathrm{def}}{=}}

\newcommand{\IC}{\mathbb{C}}
\newcommand{\IN}{\mathbb{N}}

\newcommand{\IZ}{\mathbb{Z}}
\newcommand{\ii}{\mathrm{i}}
\newcommand{\ee}{\mathrm{e}}

\newcommand{\fraksl}{\mathfrak{sl}}
\newcommand{\slh}{\widehat{\mathfrak{sl}}}
\newcommand{\frakg}{\mathfrak{g}}
\newcommand{\hfrakg}{\widehat{\mathfrak{g}}}

\newcommand{\hG}{\widehat{G}}

\newtheorem{theorem}{Theorem}
\newtheorem{proposition}{Proposition}
\newtheorem{lemma}{Lemma}

\title{Affine Lie group approach to 
a derivative nonlinear Schr\"odinger equation
and its similarity reduction}
\author{
Saburo K{\sc akei}\\
{\normalsize Department of Mathematics, Rikkyo University}\\
{\normalsize Nishi-ikebukuro, Toshima-ku, Tokyo 171-8501, Japan}\\
{\normalsize E-mail: kakei@rkmath.rikkyo.ac.jp}\\[3mm]%
Tetsuya K{\sc ikuchi}\\
{\normalsize Mathematical Institute, Tohoku University}\\
{\normalsize Aoba, Sendai 980-8578, Japan}\\
{\normalsize E-mail: tkikuchi@math.tohoku.ac.jp}}
\date{\today}

\begin{document}
\maketitle
\begin{abstract}
The generalized Drinfel'd-Sokolov hierarchies studied by 
de Groot-Hollowood-Miramontes are extended from the
viewpoint of Sato-Wilson dressing method.
In the $A_1^{(1)}$ case, we obtain the hierarchy
that include the derivative nonlinear Schr\"odinger equation.
We give two types of affine Weyl group symmetry of
the hierarchy
based on the Gauss decomposition of the $A_1^{(1)}$ affine Lie group. 
The fourth Painlev\'e equation and
their Weyl group symmetry are obtained
as a similarity reduction.
We also clarify the connection between these systems and
monodromy preserving deformations.
\end{abstract}

\tableofcontents

\section{Introduction}
There are many brilliant works on the relation between Lie algebras
and soliton equations. Among those works, the approach due to 
Drinfel'd and Sokolov \cite{DS} is a milestone, and 
gives a method for classifying many soliton equations. 
Although extended version of their work has been proposed
\cite{gds1,tau,BtK2}, there still exist several soliton equations that 
are not treated along the line of Drinfel'd-Sokolov's works. 
One example of those equations is 
a derivative nonlinear Sch\"odinger ($\pa$NLS) equation:
\begin{equation}
\label{DerNLS0}
\ii q_T = \frac{1}{2}q_{XX} + 2\ii q^2\bar{q}_X +4|q|^4q, 
\end{equation}
which has been studied by several authors \cite{ARS,GI,OS,Tsuchida}.
This integrable equation is a modification of the nonlinear Schr\"odinger 
(NLS) equation:
\begin{equation}
\ii q_T = \frac{1}{2}q_{XX} + 4|q|^2q.
\label{NLS0}
\end{equation}
Hereafter we will forget the complex structure 
of \eqref{DerNLS0}, \eqref{NLS0} and  
consider nonlinear coupled equations, 
\begin{equation}
\label{DerNLS1}
\left\{\begin{aligned}
 q_t &= \frac{1}{2}q_{xx} - 2q^2r_x -4q^3r^2, \\
 r_t &= -\frac{1}{2}r_{xx} - 2r^2q_x +4r^3q^2,
\end{aligned}\right.
\end{equation}
and
\begin{equation}
\label{NLS}
\left\{\begin{aligned}
q_{t} &= \frac{1}{2}q_{xx} + 4q^2 r, \\
r_{t} &= -\frac{1}{2}r_{xx} - 4q r^2.
\end{aligned}\right.
\end{equation}
We note that \eqref{DerNLS1}, \eqref{NLS} is reduced 
to \eqref{DerNLS0}, \eqref{NLS0}, respectively, 
under the condition $r=\bar{q}$, $X=\ii x$, $T=\ii t$.
It is well-known that 
the hierarchy of soliton equations
including NLS \eqref{NLS} is obtained as a
Drinfel'd-Sokolov hierarchy 
of $A_1^{(1)}$ homogeneous type.

The aim of the present article is threefold: 
\begin{description}
\item[Extension of the Drinfel'd-Sokolov formulation] \ \\
We extend the generalized Drinfel'd-Sokolov
hierarchy \cite{gds1} from the
viewpoint of Sato-Wilson dressing method.
The extended version includes 
the $\pa$NLS equation \eqref{DerNLS1}
as an $A_1^{(1)}$ case.

\item[Description of affine Weyl group symmetry] \ \\
There exist transformations
of the $\pa$NLS equation, called B\"acklund 
transformations 
that relate two solutions of the $\pa$NLS equation.
We construct two types of B\"acklund transformations
that satisfy the relation of the $A_1^{(1)}$ affine Weyl group. 
We remark that our construction of 
the affine Weyl group symmetry is an extention of 
the work by Noumi and Yamada \cite{NY:A, PIV}.

\item[Algebraic description of similarity reduction] \ \\
An interesting feature of the $\pa$NLS equation \eqref{DerNLS0} is 
its connection to the fourth Painlev\'e 
equation ($\mathrm{P_{IV}}$): 
\begin{equation}
 y'' = \frac{1}{2y}(y')^2 + \frac{3}{2}y^3 + 4xy^2 
+ 2(x^2 - \nu_1)y + \frac{\nu_2}{y},
\label{Painleve4}
\end{equation}
where $\nu_1, \nu_2 \in \IC$ are parameters.
In \cite{ARS}, Ablowitz, Ramani and Segur  
have shown that the self-similar solutions
of $\pa$NLS satisfy $\mathrm{P_{IV}}$ with a
special case of the parameters.
We give a systematic framework of 
similarity reductions
of the $\pa$NLS hierarchy
that gives $\mathrm{P_{IV}}$ with full-parameters. 
We also give a relation to
monodromy preserving deformation
studied by Jimbo, Miwa and Ueno \cite{JMU, JM2, JM3}.
\end{description}

As for an application to discrete integrable systems, we consider 
a discrete equation, 
\begin{equation}
 X_{n-1} + X_n + X_{n+1} 
= x + \frac{\kappa_1n + \kappa_2 + \kappa_3(-1)^n}{X_n}.
\label{dPI}
\end{equation}
This equation called the asymmetric discrete 
Painlev\'e I ($\mathrm{dP_I}$) because a 
continuous limit of \eqref{dPI} with $\kappa_3 = 0$ is the
first Painlev\'e equation. 
Grammaticos and Ramani\cite{GR}
obtained the equation \eqref{dPI}
from the Schlesinger transformations,
which are special type of B\"acklund transformations.
We construct a Schlesinger transformations of the $\pa$NLS
equation as an extension of affine Weyl group symmetry
and obtain $\mathrm{dP_I}$.

The equations, NLS \eqref{NLS}, $\mathrm{P_{IV}}$ 
\eqref{Painleve4}, and $\mathrm{dP_I}$ \eqref{dPI}, 
share a class of rational solutions 
expressed by the Hermite polynomials \cite{IY,PIV,OKS}.
We clarify the algebraic structure of this class of solutions
by using the fermionic representation of $\slh_2$.

\section{Construction of the $\pa$NLS hierarchy}
\subsection{General framework}
In this subsection, we outline our formulation of soliton equations 
based on the approach of Drinfel'd and Sokolov\cite{BtK2,gds1,DS,Wil}. 

Let $\frakg$ be a simple finite-dimensional complex Lie algebra, 
and $(\:,\:)$ be the normalized invariant scalar product of $\frakg$. 
The affine Lie algebra $\hfrakg$ associated to $(\frakg,\,(\:,\:))$ 
can be realized as
\begin{equation*}
 \hfrakg = \frakg\otimes\IC[z,z^{-1}]\oplus\IC K\oplus\IC d, 
\end{equation*}
with the relations, 
\begin{equation*}
\begin{array}{l}
[X\otimes z^m, Y\otimes z^n]=[X,Y]\otimes
 z^{m+n}+m\delta_{m+n,0}(X,Y)K,\\
{[K,\hfrakg]}=0,\quad [d, X\otimes z^n]=nX\otimes z^n, 
\end{array}
\end{equation*}
for $X,Y\in\hfrakg$, $m,n\in\IZ$ \cite{Kacbook}. 

To construct integrable hierarchies, Heisenberg subalgebras of 
$\hfrakg$ play a crucial role. It is known that non-equivalent 
Heisenberg subalgebras are classified by conjugacy classes of the 
Weyl group of $\frakg$ \cite{112,gds1}. 
We denote  by $\mathcal{H}^{[w]}$
the Heisenberg subalgebra 
associated with the conjugacy class $[w]$:
$$
  \mathcal{H}^{[w]} = \bigoplus_{n\in\IZ} \IC \Lambda_n^{[w]}
                        \oplus \IC K.
$$ 

Once we fix a basis of Heisenberg subalgebra $\{\Lambda_n^{[w]}\}_{n\in\IZ}$, 
there is an associated gradation $d_w$ that is natural 
on $\{\Lambda_n^{[w]}\}_{n\in\IZ}$: 
\begin{equation*}
 [d_w, \Lambda_n^{[w]}] = n\Lambda_n^{[w]}. 
\end{equation*}
The gradation $d_w$ induces a $\IZ$-grading on $\hfrakg$: 
\begin{equation*}
 \hfrakg = \mathop{\bigoplus}_{j\in\IZ}\hfrakg_j^{[w]}, \quad 
 \hfrakg_j^{[w]}=\{x\in\hfrakg\; ; \;[d_w,x]=jx\}.
\end{equation*}
For an integer $k$, we use the notation 
\begin{equation*}
 \hfrakg_{\ge k}^{[w]}= 
\mathop{\bigoplus}_{j\ge k}\hfrakg_j^{[w]}, \quad
 \hfrakg_{<k}^{[w]}= 
\mathop{\bigoplus}_{j<k}\hfrakg_j^{[w]}.
\end{equation*}

We now consider a Kac-Moody group $\hG$ formed by exponentiating 
the action of $\hfrakg$ on a integrable module. 
Throughout this paper, we assume that the exponentiated action of 
an element of the positive degree subalgebra
of $\mathcal{H}^{[w]}$ is well-defined. 
We remark that all of the representations used in what follows 
belong to this category.
We denote by $\hG^{[w]}_{\ge 0}$ and 
$\hG^{[w]}_{<0}$ the subgroups correspond to the subalgebras 
$\hfrakg_{\ge 0}^{[w]}$ and $\hfrakg_{<0}^{[w]}$, respectively. 

Starting from $g(0) \in \hG$, we define time-evolutions 
with time variable $t=(t_1, t_2, \dots)$ using 
the Heisenberg subalgebra $\{\Lambda_n^{[w']}\}_{n\in\IZ}$ 
associated with a conjugacy class $[w']$: 
\begin{equation}
\label{defg(t)}
 g(t)\defeq \exp\left(\sum_{n>0}t_n\Lambda_n^{[w']}\right)g(0), 
\end{equation}
which satisfies the following differential equation, 
\begin{equation}
\label{LinearDE}
 \frac{\pa g(t)}{\pa t_n}=\Lambda_n^{[w']}g(t), \quad n=1,2,\ldots
\end{equation}
In what follows, we shall assume the existence and the uniqueness 
of the Gauss decomposition
with respect to the gradation $d_w$: 
\begin{equation}
\label{GaussDecomp}
g(t) = \{g^{[w]}_{<0}(t) \}^{-1}g^{[w]}_{\ge 0}(t), \qquad
g^{[w]}_{<0}(t) \in \hG^{[w]}_{<0}, \quad
g^{[w]}_{\ge 0}(t) \in \hG^{[w]}_{\ge 0}. 
\end{equation}
A detailed discussion about this assumption is 
in \cite{BtK1, Wil} for instance. 
Note that the conjugacy classes of Weyl 
group $[w]$ of \eqref{GaussDecomp}
and $[w']$ of \eqref{defg(t)} is not necessary equal. 
From \eqref{LinearDE} and \eqref{GaussDecomp}, 
we have 
\begin{align}
\frac{\pa g^{[w]}_{<0}}{\pa t_n} &= 
   B_n g^{[w]}_{<0}-g^{[w]}_{<0}\Lambda_n^{[w']}, 
\label{SatoEq<} 
\\
\frac{\pa g^{[w]}_{\ge 0}}{\pa t_n} &= B_n g^{[w]}_{\ge 0},
\label{SatoEq>=}
\end{align}
where $B_n=B_n(t)$ is defined by 
\begin{equation}
B_n(t)\defeq\left(g^{[w]}_{<0}(t)
\Lambda_n^{[w']}g^{[w]}_{<0}(t)^{-1}\right)_{\ge 0}^{[w]}
\in\hfrakg_{\ge 0}^{[w]}.
\label{defofBn}
\end{equation}
We call \eqref{SatoEq<} and \eqref{SatoEq>=} 
the Sato-Wilson equations.
The compatibility conditions for \eqref{SatoEq<} or \eqref{SatoEq>=} 
give rise to the zero-curvature (or Zakharov-Shabat) equations, 
\begin{equation}
\label{ZakSab}
\frac{\pa B_m}{\pa t_n}-\frac{\pa B_n}{\pa t_m}
+[B_m,B_n]=0, \quad m,n=1,2,\ldots, 
\end{equation}
which gives a hierarchy of soliton equations.

Note that de Groot et al.~imposed 
the condition $w' = w$ and in the definition 
\eqref{defofBn}
of $B_n$, they used a projection 
with respect to $d_w$ or a less gradation than $d_w$
for an order of gradations \cite{gds1}.
However the formulas \eqref{SatoEq<} and \eqref{SatoEq>=}
is valid without the relation for $w$ and $w'$.
In this sense, our formulation can
be regarded as an extension of the generalized 
Drinfel'd-Sokolov hierarchy.

\subsection{Hierarchy of the derivative NLS equation}
Hereafter we consider only the $\slh_2$-case to treat the 
$\pa$NLS hierarchy. 
The generators of $\fraksl_2$ is denoted by $E$, $F$ and $H$ as usual: 
\begin{equation*}
 [E, F] =H, \quad [H, E]=2E, \quad [H, F]=-2F. 
\end{equation*}
We will use the abbreviation $X_n=X\otimes z^n$ for $X=E,F,H$. 

In the case of $\fraksl_2$, corresponding Weyl group is the 
symmetric group $\mathfrak{S}_2$ of order 2, generated by 
the simple transposition $\sigma$. 
The gradation corresponds to Id$\in\mathfrak{S}_2$ is given by 
the element $d$, and called ``homogeneous''.
The gradation corresponds to $\sigma\in\mathfrak{S}_2$ is called 
``principal'', given by $d_\mathrm{p}=2d+\frac{1}{2}H_0$.

We choose the Heisenberg subalgebra 
of homogeneous-type, 
\begin{equation}
\label{homoHeisenberg}
 \Lambda_n^{[\mathrm{h}]}\defeq H_n, 
 \end{equation}
and the triangular decomposition of principal-type, 
\begin{equation*}
\slh_2 = \left(\slh_2\right)_{< 0}^{[\mathrm{p}]}
\oplus \left(\slh_2\right)_{\ge 0}^{[\mathrm{p}]}. 
\end{equation*}
In other words, we have chosen $w'=\mathrm{Id}$ in \eqref{defg(t)} 
and $w=\sigma$ in \eqref{GaussDecomp}. We stress that this choice 
does not fit to the condition $w'\ge w$ in the 
sense of the Bruhat order, and thus does not fall into the 
category treated in \cite{gds1}. 
Note that the homogeneous Heisenberg 
subalgebra \eqref{homoHeisenberg} has even principal grades: 
\begin{equation*}
[d_\mathrm{p},\Lambda_n^{[\mathrm{h}]}]=2n\Lambda_n^{[\mathrm{h}]}.
\end{equation*}

We consider a formal series expansion of 
$\log g^{[\mathrm{p}]}_{<0}(t) \in \slh_2$ as follows: 
\begin{align}
\log g^{[\mathrm{p}]}_{<0}(t) 
&= \{q(t)E_{-1}+r(t)F_0\}+ u(t)H_{-1}\nonumber\\
&\quad + \{v_1(t)E_{-2}+v_2(t)F_{-1}\}+ w(t)H_{-2}+\cdots.
\label{formalExpansion}
\end{align}
By straightforward calculations, we can obtain the expression 
for $B_1$: 
\begin{equation}
B_1 = H_1 + (-2q E_0 +2r F_1)+ \left\{2qr H_0-(qr+2u)K\right\}.
\label{B_1}
\end{equation}
\begin{lemma}
\label{lemma:reduction}
\begin{equation}
\label{lamma1}
\frac{\pa q}{\pa t_1}=2v_1+2qu+\frac{4}{3}q^2r,\quad
\frac{\pa r}{\pa t_1}=-2v_2+2ru-\frac{4}{3}qr^2
\end{equation}
\end{lemma}
\begin{proof}
In the present case, the Sato-Wilson 
equation \eqref{SatoEq<} with $n=1$ is equivalent to 
the following equation in $\slh_2$:
\begin{equation}
\label{SatoEq<'}
 \frac{\pa g^{[\mathrm{p}]}_{<0}}{\pa t_1}\;
       (g^{[\mathrm{p}]}_{<0})^{-1} 
= B_1 - g^{[\mathrm{p}]}_{<0} \; H_1 (g^{[\mathrm{p}]}_{<0})^{-1}.
\end{equation}
Comparing the $(\:\cdot\:)_{-1}^{[\mathrm{p}]}$-part of the 
both side of \eqref{SatoEq<'}, we can derive the desirous result.
\end{proof}
This lemma gives us the expression for
$B_2$: 
\begin{align}
B_2 &= 
H_2 + (-2qE_1+2rF_2) + 2qr H_1
+ \left(-q'E_0-r'F_1\right)\nonumber\\
&\qquad + \left\{ \left(q'r-qr'-2q^2r^2 \right)H_0
+\left(-4w-rv_1-\frac{2}{3}qru-\frac{1}{3}q^2r^2\right)K
\right\}.\label{B_2}
\end{align}
Here and throughout this paper, $\prime$ denotes 
partial differentiation with respect to $t_1$. 
Substituting these expressions into \eqref{ZakSab}, 
we can obtain the $\pa$NLS equation \eqref{DerNLS1} for $x = t_1, t = t_2$. 
In this sense, the hierarchy now we consider is nothing but 
the $\pa$NLS hierarchy.

A level-$0$ realization of $\slh_2$ is given by
\begin{align}
&E_n \mapsto \begin{pmatrix}0 & z^n\\ 0 & 0\end{pmatrix}, \quad
F_n \mapsto \begin{pmatrix}0 & 0\\ z^n & 0\end{pmatrix}, \quad
H_n \mapsto \begin{pmatrix}z^n & 0\\ 0 & -z^n\end{pmatrix}, 
\nonumber \\
\label{level0realization1}
&K \mapsto 0,  \quad
 d \mapsto z\frac{d}{dz}.
\end{align}
Using this realization, we can express $B_1$ and $B_2$ as 
$2\times 2$ matrices: 
\begin{align*}
\bm{B}_1 &= 
\begin{pmatrix}z & 0 \\ 0 & -z\end{pmatrix}
+ \begin{pmatrix}0 & -2q\\ 2zr & 0\end{pmatrix}
+ \begin{pmatrix}2qr & 0 \\ 0 & -2qr\end{pmatrix},\\
\bm{B}_2 &=
\begin{pmatrix}z^2 & 0 \\ 0 & -z^2\end{pmatrix}
+ \begin{pmatrix}0 & -2zq\\ 2z^2r & 0\end{pmatrix}
+ \begin{pmatrix}2zqr & 0 \\ 0 & -2zqr\end{pmatrix}
\nonumber\\
&\quad  + \begin{pmatrix}0 & -q'\\ -zr' & 0\end{pmatrix}
+ \begin{pmatrix}q'r-qr'-2q^2r^2 & 0 \\
0 & qr'-q'r+2q^2r^2\end{pmatrix}.
\end{align*}
These matrices give a Lax pair for the $\pa$NLS equation 
but are different from the conventional one (cf.~\cite{WS}). 
To reproduce the conventional Lax pair, we use 
the other level-$0$ realization of $\slh_2$ given by 
\begin{align}
&E_n \mapsto \begin{pmatrix}0 & \lambda^{2n+1}\\ 0 & 0\end{pmatrix}, \quad
F_n \mapsto \begin{pmatrix}0 & 0\\ \lambda^{2n-1} & 0\end{pmatrix}, \quad
H_n \mapsto \begin{pmatrix}\lambda^{2n} & 0\\
 0 & -\lambda^{2n}\end{pmatrix}, 
\nonumber \\
&K \mapsto 0,  \quad
d \mapsto \frac{1}{2}\left\{z\frac{d}{dz} 
+ \frac{1}{2}\begin{pmatrix}1 & 0 \\ 0 & -1 \end{pmatrix}\right\}. 
\label{real2}
\end{align}
{}From this realization, we obtain 
\begin{align*}
\bm{B}_1 &= 
 \lambda^2\begin{pmatrix}1 & 0 \\ 0 & -1\end{pmatrix}
+\lambda\begin{pmatrix}0 & -2q\\ 2r & 0\end{pmatrix}
+\begin{pmatrix}2qr & 0 \\ 0 & -2qr\end{pmatrix},\\
\bm{B}_2 &= 
\lambda^4\begin{pmatrix}1 & 0 \\ 0 & -1\end{pmatrix}
+\lambda^3\begin{pmatrix}0 & -2q\\ 2r & 0\end{pmatrix}
+\lambda^2\begin{pmatrix}2qr & 0 \\ 0 & -2qr\end{pmatrix}\nonumber\\
& \quad + 
\lambda\begin{pmatrix}0 & -q'\\ -r' & 0\end{pmatrix}
+\begin{pmatrix}q'r-qr'-2q^2r^2 & 0 \\
0 & qr'-q'r+2q^2r^2\end{pmatrix}.
\end{align*}

For the latter use, we decompose $g^{[\mathrm{p}]}_{\ge 0}(t)$ into 
grade $0$ and $>0$ part: 
\begin{equation}
g^{[\mathrm{p}]}_{\ge 0}(t) = 
g^{[\mathrm{p}]}_{0}(t)g^{[\mathrm{p}]}_{>0}(t).
\label{Decom0and>0}
\end{equation}
Substituting \eqref{Decom0and>0} into \eqref{SatoEq>=}, we obtain 
\begin{align*}
\frac{\pa g^{[\mathrm{p}]}_0}{\pa t_n}
&=\left((g^{[\mathrm{p}]}_0)^{-1}B_ng^{[\mathrm{p}]}_0
\right)^{[\mathrm{p}]}_0 g^{[\mathrm{p}]}_0, \\
\frac{\pa g^{[\mathrm{p}]}_{>0}}{\pa t_n}
&= \left\{(g^{[\mathrm{p}]}_0)^{-1}B_ng^{[\mathrm{p}]}_0
-\left((g^{[\mathrm{p}]}_0)^{-1}B_ng^{[\mathrm{p}]}_0
\right)^{[\mathrm{p}]}_0 \right\}
g^{[\mathrm{p}]}_{>0}. 
\end{align*}
{}From these differential equations together with 
\eqref{B_1}, \eqref{B_2} and formal expansions, 
\begin{align}
\log g^{[\mathrm{p}]}_{0}(t) &= \phi(t)H_0 + \psi(t)K,
\label{expansion_=0} \\
\log g^{[\mathrm{p}]}_{>0}(t) &= 
  a(t)E_0 + b(t)F_1 + c(t)H_1 + \cdots, 
\label{expansion_>0}
\end{align}
it follows that the functions $\phi(t)$, $a(t)$, $b(t)$ 
satisfy the equations, 
\begin{alignat}{2}
  \frac{\pa\phi}{\pa t_1}&= 2qr, 
 &\frac{\pa\phi}{\pa t_2}&= q'r-qr'-2q^2r^2, 
\label{diffEq_for_phi} \\
  \frac{\pa a}{\pa t_1}&= -2q\,\ee^{-2\phi}, \qquad
 &\frac{\pa a}{\pa t_2}&= -q'\,\ee^{-2\phi}, 
\label{diffEq_for_a} \\
  \frac{\pa b}{\pa t_1}&= 2r\,\ee^{2\phi},
 &\frac{\pa b}{\pa t_2}&= -r'\,\ee^{2\phi}. 
\label{diffEq_for_b}
\end{alignat}

\subsection{Miura-type transformation to the NLS equation}

The homogeneous hierarchy that includes
the NLS equation is obtained by 
taking $w' = \mathrm{Id}$ in the 
time-evolution \eqref{defg(t)},
same as $\pa$NLS hierarchy,
and $w = \mathrm{Id}$ in the Gauss decomposition \eqref{GaussDecomp}.
We denote the result of the decomposition by
$$
 g(t) = \{ g_{<0}^{[\mathrm{h}]}(t) \}^{-1} g_{\ge 0}^{[\mathrm{h}]}(t), 
\qquad
 g_{<0}^{[\mathrm{h}]}(t) \in \hG_{<0}^{[\mathrm{Id}]}, \quad
 g_{<0}^{[\mathrm{h}]}(t) \in \hG_{\ge 0}^{[\mathrm{Id}]}. 
$$
The relation between this system and the $\partial$NLS hierarchy is
established by the Miura-type transformation, 
which is an analog of the Miura transformation in
the case of the KdV and the mKdV equations.
For $g^{[\mathrm{p}]}_{<0}(t)$ of \eqref{formalExpansion}, 
we put 
\begin{equation*}
   G \defeq \exp(-r(t)F_0)
\end{equation*}
and consider the decomposition,
$$
 g(t) = \{ Gg_{<0}^{[\mathrm{p}]}(t) \}^{-1}
         G g_{\ge 0}^{[\mathrm{p}]},
\qquad
 Gg^{[\mathrm{p}]}_{<0}(t) \in \hG_{<0}^{[\mathrm{Id}]}, \quad
 Gg^{[\mathrm{p}]}_{\ge 0}(t) \in \hG_{\ge 0}^{[\mathrm{Id}]}.
$$
The assumption of the uniqueness of the Gauss decomposition causes
\begin{align}
 g^{[\mathrm{h}]}_{<0}(t) 
&= Gg^{[\mathrm{p}]}_{<0}(t) = \exp(qE_{-1} + uH_{-1} + \cdots)  
\in \hG_{<0}^{[\mathrm{Id}]}, 
\label{homog<0} \\
 g^{[\mathrm{h}]}_{\ge 0}(t) &= Gg^{[\mathrm{p}]}_{\ge 0}(t) 
\in \hG_{\ge 0}^{[\mathrm{Id}]}.
\label{homog>=0}
\end{align}
These relations can be considered as
Miura-type transformation in affine Lie group.
By the equation \eqref{SatoEq<}, we can write
\begin{equation}
 \frac{\partial}{\partial t_n} - B_n
= g^{[\mathrm{p}]}_{< 0} \left( \frac{\partial}{\partial t_n}
- H_n \right) (g^{[\mathrm{p}]}_{<0})^{-1}.
\label{SatoEq<2}
\end{equation}
Then we can describe the transformation in terms 
of $B_n$ by translating $g^{[\mathrm{p}]}_{<0}(t)$ to 
$g^{[\mathrm{h}]}_{<0}(t)$ in \eqref{SatoEq<2} and we obtain
$$
  \frac{\partial}{\partial t_n} - \tilde{B}_n
\defeq  g^{[\mathrm{h}]}_{< 0} \left( \frac{\partial}{\partial t_n}
- H_n \right) (g^{[\mathrm{h}]}_{<0})^{-1}
= G \left( \frac{\partial}{\partial t_n}
- B_n \right) G^{-1}.
$$
Note that this transformation preserves the zero-curvature 
equations \eqref{ZakSab}. 
The  relation between $\tilde{B}_n$ and $B_n$ can be
described as follows:
\begin{equation}
 \tilde{B}_n = G B_n G^{-1} + 
\frac{\partial G}{\partial t_n}G^{-1}.
\label{Btilde}
\end{equation}
For $n=1, 2$, we obtain
\begin{align}
 \tilde{B}_1 =& H_1 + (- 2qE_0 - (r' + 2qr^2)F_0)
                 - (qr + 2u)K,
\label{B_1tilde} \\
 \tilde{B}_2 =& H_2 + (-2qE_1 - (r' + 2qr^2)F_1)
- q(r' + 2qr^2)H_0 \nonumber \\
& \quad -q'E_0 + \left(\frac{r'}{2} + qr^2 \right)'F_0
+ \left(-4w - rv_1 - \frac{2}{3}qru - \frac{1}{3}q^2r^2\right)K.
\label{B_2tilde}
\end{align}
Here we have used the $\partial$NLS equation \eqref{DerNLS1} to 
eliminate $r_t$ in $\tilde{B}_2$.

If we put
\begin{equation}
 \hat{r} = -\frac{r'}{2} - qr^2,
\label{rhat}
\end{equation}
then the zero-curvature equation for $\tilde{B}_1$ and $\tilde{B}_2$ 
gives the NLS equation \eqref{NLS}.

\subsection{Gauge transformation to a generalized 
$\pa$NLS equation}

There are several different kind of derivative NLS equations
\cite{ARS,CLL,GI,KSS,KN,Kundu}.
By extending the approach of \cite{WS},
Kundu obtained the generalized $\pa$NLS equation \cite{Kundu},
\begin{equation}
\label{GDNLS}
\left\{\begin{aligned}
 Q_t &= \frac{1}{2}Q'' + 2c QRQ' + 2(c - 1)Q^2R'
         - 2(c-1)(c - 2)Q^3R^2,  \\
 R_t &= -\frac{1}{2}R'' + 2c QRR' + 2(c - 1)R^2Q'
         + 2(c-1)(c - 2)Q^2R^3.
\end{aligned}\right.
\end{equation}
Here $c$ is a complex parameter.
The equation \eqref{GDNLS} include 
the Kaup-Newell equation $(c=1)$ \cite{KN},
the Chen-Lee-Liu equation $(c=2)$ \cite{CLL}
and also \eqref{DerNLS1} $(c=0)$ as special cases.
We can obtain the equation \eqref{GDNLS}
by the gauge transformation
of type \eqref{Btilde}
with respect to $g^{[\mathrm{p}]}_0(t)^{-c/2} 
= \exp(-(c \phi/2)H_0)$:
\begin{align*}
 \frac{\partial}{\partial t_n} - B_n
\mapsto& \; g^{[\mathrm{p}]}_0(t)^{-c/2} \left( \frac{\partial}{\partial t_n}
- B_n \right) g^{[\mathrm{p}]}_0(t)^{c/2}\\
&= \frac{\partial}{\partial t_n} 
-g^{[\mathrm{p}]}_0(t)^{-c/2}B_n g^{[\mathrm{p}]}_0(t)^{c/2} + 
g^{[\mathrm{p}]}_0(t)^{-c/2} 
\frac{\partial g^{[\mathrm{p}]}_0(t)^{c/2}}{\partial t_n}
\end{align*}
and put
$$ 
 C_n \defeq
g^{[\mathrm{p}]}_0(t)^{-c/2} B_n 
g^{[\mathrm{p}]}_0(t)^{c/2} - 
g^{[\mathrm{p}]}_0(t)^{-c/2} 
\frac{\partial g^{[\mathrm{p}]}_0(t)^{c/2}}{\partial t_n}.
$$ 
Then for $n=1, 2$, we have
\begin{align*}
 C_1 =& H_1 + (-2qe^{-c\phi}E_0 + 2re^{c\phi}F_1)
         -(c-2)qrH_0 - (qr + 2u)K\\
 C_2 =& H_2 + (-2qe^{-c\phi}E_1 + 2re^{c\phi}F_2)
         +2qrH_1  \\
      &\quad + (-q'e^{-c\phi}E_0 
            -r'e^{c\phi}F_1)
        + (1-c/2)(q'r-qr'-2q^2r^2)H_0 \\
    & \quad + \left(-4w - rv_1 - \frac{2}{3}qru - \frac{1}{3}q^2r^2\right)K.
\end{align*}
Here we have used the relation \eqref{diffEq_for_phi}.
We introduce the new variables
$$
 Q(t) \defeq qe^{-c\phi}, \qquad
 R(t) \defeq re^{c\phi}.
$$
By \eqref{diffEq_for_phi},
the derivatives of these functions are
written as
\begin{equation*}
 Q' = q'e^{-c\phi}
- 2c Q^2R,
\quad
 R' = r'e^{c\phi} + 2c R^2Q.
\end{equation*}
Then the zero-curvature 
equation for $C_1$ and $C_2$ result in the equations \eqref{GDNLS}.
Especially, 
the Lax operators $C_1$, $C_2$ for the Kaup-Newell equation and the
Chen-Lee-Liu equation
realized as matrix form \eqref{real2}
are identified with 
that of \cite{WS}.

\section{Actions of affine Weyl group to the $\pa$NLS hierarchy}
In this section, we discuss symmetries of the $\pa$NLS hierarchy 
in terms of the affine Weyl group 
$W(A^{(1)}_1)=\langle s_0,s_1\rangle$ with 
the relations $s_0^2=s_1^2=\mathrm{Id}$. 

Let $V$ be an integrable module of $\slh_2$. 
The affine Weyl group $W(A^{(1)}_1)$ 
acts on $V$ as follows \cite{Kacbook}: 
\begin{equation}
\label{WeylGenerators}
 s_j = \exp(f_j)\exp(-e_j)\exp(f_j) \quad (j=0,1),
\end{equation}
where $e_j$, $f_j$ are the Chevalley generators of $\slh_2$ 
given by 
\begin{equation*}
\begin{aligned}
e_0&=F_1, \quad&f_0&=E_{-1}, \quad&h_0&=K-H_0, \\
e_1&=E_0, &f_1&=F_0, &h_1&=H_0.
\end{aligned}
\end{equation*}
Note that $e_j, h_j, f_j$ $(j=0,1)$ have
principal grades $1, 0, -1,$ respectively.
Under the level-$0$ realization \eqref{level0realization1},
we can describe them as follows:
\begin{equation}
s_0 \mapsto \begin{pmatrix}0 & z^{-1} \\ -z & 0 \end{pmatrix},
\quad
s_1 \mapsto \begin{pmatrix}0 & -1 \\ 1 & 0 \end{pmatrix}.
\label{realization-of-Weyl}
\end{equation}
The generators $s_0$, $s_1$ act naturally on $g(0)$ of 
\eqref{defg(t)} in two different ways.

\subsection{Left-action}
We consider the left-action of $s_j$ ($j=0,1$) of the form 
$s_j^{-1}g(0)$. 
Applying the principal Gauss decomposition \eqref{GaussDecomp} to 
$\exp\!\left[\sum_n t_nH_n\right]s_j^{-1}g(0)$, we define 
$s_j^\mathrm{L}(g_{<0}(t))$ and $s_j^\mathrm{L}(g_{\ge 0}(t))$ as 
\begin{equation}
\label{GaussLeftaction}
\left\{s_j^\mathrm{L}(g_{<0}(t))\right\}^{-1}
s_j^\mathrm{L}(g_{\ge 0}(t)) = \exp\!\left[\sum_{n>0}
 t_nH_n\right]s_j^{-1}g(0). 
\end{equation}
This decomposition induces an action of $s_j$ on the 
variables $q(t)$, $r(t)$.  

\begin{theorem}
Assume that the Gauss decomposition 
\eqref{GaussLeftaction} exists uniquely. Then one 
can write down the action of $s_j$ $(j=0,1)$ explicitly: 
\begin{gather}
\label{LeftactionToqr0}
s_0^\mathrm{L} : 
q(t) \mapsto -\frac{1}{q(-t)},\quad
r(t) \mapsto q(-t)^2r(-t)-\frac{1}{2}q'(-t),\\
\label{LeftactionToqr1}
s_1^\mathrm{L} : q(t)\mapsto q(-t)r(-t)^2+\frac{1}{2}r'(-t),\quad
r(t) \mapsto -\frac{1}{r(-t)}.
\end{gather}
Here $-t = (-t_1, -t_2, \dots)$.
\end{theorem}
\begin{proof}
Using the relation $s_jH_ns_j^{-1}=-H_n$ $( j=0,1$, $n=1,2,\ldots )$, 
one can rewrite \eqref{GaussLeftaction} as 
\begin{equation*}
\left\{s_j^\mathrm{L}(g_{<0}(t))\right\}^{-1}
s_j^\mathrm{L}(g_{\ge 0}(t)) = 
s_j^{-1}g(-t) = 
\left\{g^{[\mathrm{p}]}_{<0}(-t)s_j\right\}^{-1}
g^{[\mathrm{p}]}_{\ge 0}(-t).
\end{equation*}

Next we consider the Gauss decomposition of 
$\left\{g^{[\mathrm{p}]}_{<0}(-t)s_j\right\}^{-1}$: 
\begin{equation}
\label{decomp_of_g<}
\left\{g^{[\mathrm{p}]}_{<0}(-t)s_j\right\}^{-1} 
= \left\{
\check{g}_{<0}^{(j)}(t)\right\}^{-1}\check{g}^{(j)}_{\ge 0}(t). 
\end{equation}
Assuming the uniqueness of the Gauss decomposition 
\eqref{GaussLeftaction}, one finds that 
$$
 s_j^\mathrm{L}(g_{<0}(t)) = \check{g}^{(j)}_{<0}(t).
$$ 
The decomposition \eqref{decomp_of_g<} is equivalent to 
the condition, 
\begin{equation}
\label{decomp_of_g<_2}
g^{[\mathrm{p}]}_{<0}(-t)s_j
\left\{\check{g}^{(j)}_{<0}(t)\right\}^{-1}
\in \hG^{[\mathrm{p}]}_{\ge 0}.
\end{equation}

We introduce a formal series expansion of 
$\log\check{g}^{(j)}_{<0}(t)$ as 
\begin{align}
\log\check{g}^{(j)}_{<0}(t)
&= \{\check{q}_j(t)E_{-1}+\check{r}_j(t)F_0\}
+ \check{u}_j(t)H_{-1}\nonumber\\
&\quad + \{\check{v}_{1,j}(t)E_{-2}+\check{v}_{2,j}(t)F_{-1}\}
+ \check{w}_j(t)H_{-2}+\cdots. 
\label{LeftactionToqr}
\end{align}
Substituting $g^{[\mathrm{p}]}_{<0}(-t)$ \eqref{formalExpansion}, 
$s_j$ \eqref{WeylGenerators} and
\eqref{LeftactionToqr} into 
$g^{[\mathrm{p}]}_{<0}(-t)s_j\left\{\check{g}^{(j)}_{<0}(t)\right\}^{-1}$ 
and using the realization \eqref{level0realization1}, 
we can rewrite the condition \eqref{decomp_of_g<_2} as 
relations between the coefficients of 
$g^{[\mathrm{p}]}_{<0}(-t)$ \eqref{formalExpansion} 
and $\check{g}^{[\mathrm{p}]}_{<0}(t)$ \eqref{LeftactionToqr}. 
For example, in the case of $s_1$, we have 
\begin{align*}
& 1+r(-t)\check{r}_1(t)=0, \\
& u(-t)+\check{u}_1(t)
 +\frac{q(-t)r(-t)+\check{q}_1(t)\check{r}_1(t)}{2}=0, \\
& v_2(-t)+\frac{q(-t)r(-t)^2}{6} + \check{q}_1(t) +
  \frac{r(-t)\check{q}_1(t)\check{r}_1(t)}{2} + r(-t)\check{u}_1(t) =0. 
\end{align*}
{}From these relations together with \eqref{lamma1}, 
we obtain \eqref{LeftactionToqr1}. 
The $s_0$-action \eqref{LeftactionToqr0} can be obtained in a similar way.
\end{proof}

We remark that $-s_1^\mathrm{L}(q(-t))$ coincides with $\hat{r}$ 
of \eqref{rhat}. 
Thus $q(t)$ and $\hat{r}(t) = -s_1^\mathrm{L}(q(-t))$ solve 
the NLS equation \eqref{NLS}.

\subsection{Right-action}
Next we consider the right-action of $s_j$ ($j=0,1$) of the form 
$g(0)s_j$, which induces another action of $s_j$ on the 
variables $q(t)$, $r(t)$ through the decomposition, 
\begin{equation}
\label{GaussRightaction}
\left\{s_j^\mathrm{R}(g^{[\mathrm{p}]}_{<0}(t))\right\}^{-1}
s_j^\mathrm{R}(g^{[\mathrm{p}]}_{\ge 0}(t)) = g(t)s_j. 
\end{equation}
\begin{theorem}
Assume that the Gauss decomposition 
\eqref{GaussRightaction} exists uniquely. Then one 
can write down the action of $s_j$ $(j=0,1)$ explicitly: 
\begin{gather}
\label{RightactionToqr0}
s_0^\mathrm{R} : 
q(t) \mapsto q(t)-\frac{1}{\tilde{\psi}_0(t)}, \quad
r(t) \mapsto r(t),\\
\label{RightactionToqr1}
s_1^\mathrm{R} : q(t) \mapsto q(t),\quad
r(t) \mapsto r(t)-\frac{1}{\tilde{\psi}_1(t)}. 
\end{gather}
Here $\tilde{\psi}_0(t)$ and $\tilde{\psi}_1(t)$ satisfy the 
following differential equations, 
\begin{alignat}{2}
\frac{\pa\tilde{\psi}_0}{\pa t_1}&=2r-4qr\tilde{\psi}_0, 
&\frac{\pa\tilde{\psi}_0}{\pa t_2}
&= -r' - 2(q'r - qr' - 2q^2r^2)\tilde{\psi}_0,
\label{diffEq_for_f0}\\
\frac{\pa\tilde{\psi}_1}{\pa t_1}&=-2q+4qr\tilde{\psi}_1, 
\qquad
&\frac{\pa\tilde{\psi}_1}{\pa t_2}
& = -q' + 2(q'r - qr' - 2q^2r^2)\tilde{\psi}_1.
\label{diffEq_for_f1}
\end{alignat}
\end{theorem}
\begin{proof}
We consider the Gauss decomposition of 
$g^{[\mathrm{p}]}_{\ge 0}(t)s_j$: 
\begin{equation}
\label{decomp_of_g>=}
g^{[\mathrm{p}]}_{\ge 0}(t)s_j= \left\{
\tilde{g}^{(j)}_{<0}(t)\right\}^{-1}
\tilde{g}^{(j)}_{\ge 0}(t). 
\end{equation}
Assuming the uniqueness of the Gauss decomposition 
\eqref{GaussRightaction}, one finds that 
\begin{equation}
  s_j^\mathrm{R}(g^{[\mathrm{p}]}_{<0}(t)) 
= \tilde{g}^{(j)}_{<0}(t)g^{[\mathrm{p}]}_{<0}(t).
\label{sjR}
\end{equation}
The decomposition \eqref{decomp_of_g>=} is equivalent to 
\begin{equation}
\label{decomp_of_g>=_2}
 g^{[\mathrm{p}]}_{\ge 0}(t)s_j
\left\{\tilde{g}^{(j)}_{\ge 0}(t)\right\}^{-1}
=\left\{ \tilde{g}^{(j)}_{<0}(t) \right\}^{-1}
\in \hG^{[\mathrm{p}]}_{<0}.
\end{equation}
In the realization \eqref{level0realization1}, 
substituting formal expansions $g^{[\mathrm{p}]}_{\ge 0}(t) = 
g^{[\mathrm{p}]}_{0}(t)g^{[\mathrm{p}]}_{>0}(t)$ \eqref{expansion_=0}, 
\eqref{expansion_>0},
$\tilde{g}^{(j)}_{\ge 0}(t) = 
\tilde{g}^{(j)}_{0}(t)\tilde{g}^{(j)}_{>0}(t)$ 
$$
\log \tilde{g}^{(j)}_{0}(t) = \tilde{\phi}_j(t)H_0,
\quad
\log\tilde{g}^{(j)}_{> 0}(t) = 
\tilde{a}_j(t)E_0+\tilde{b}_j(t)F_1+\tilde{c}_j(t)H_1+\cdots, 
$$
and $s_j$ \eqref{WeylGenerators}
to \eqref{decomp_of_g>=_2}, 
we have
\begin{align*}
 &b\ee^{\tilde{\phi}_0 - \phi} =1, \quad 
  b\tilde{b}_0 = -1, \quad \dots \\ 
 &a\ee^{\phi - \tilde{\phi}_1} =1, \quad 
  a\tilde{a}_1 = -1, \quad \dots 
\end{align*}
and
$$
\tilde{g}^{(0)}_{<0}(t) = \exp[-\ee^{\phi + \tilde{\phi}_0}E_{-1}], 
\qquad 
\tilde{g}^{(1)}_{<0}(t) = \exp[-\ee^{-\phi - \tilde{\phi}_1}F_0].
$$
Therefore,
we obtain $s_j^\mathrm{R}(g^{[\mathrm{p}]}_{<0}(t))$ $(j=0,1)$
from \eqref{sjR}.
If we define 
\begin{equation}
 \tilde{\psi}_0 = \ee^{-\phi-\tilde{\phi}_0} = \ee^{-2\phi}b, \qquad
 \tilde{\psi}_1 = \ee^{\phi+\tilde{\phi}_1} = \ee^{2\phi}a
\label{ftilde}
\end{equation}
we have the formulas 
\eqref{RightactionToqr0}, \eqref{RightactionToqr1}. 
The differential equations \eqref{diffEq_for_f0},
\eqref{diffEq_for_f1} follow from 
\eqref{diffEq_for_phi}, \eqref{diffEq_for_a} and \eqref{diffEq_for_b}.
\end{proof}
We remark that the right action can be described by the gauge
transformation of the differential operators:
$$
  \frac{\partial}{\partial t_n} - s_j(B_n)
= \tilde{g}^{(j)}_{< 0} 
  \left( \frac{\partial}{\partial t_n} - B_n \right) 
  (\tilde{g}^{(j)}_{<0})^{-1}
\qquad
(j=0,1).
$$
This construction of the Weyl group action is essentially 
the same as that of Noumi and Yamada \cite{NY:A, ICM}.
We will discuss this point in what follows
(See Section \ref{SymP4}).

\subsection{Extended affine Weyl group}
We denote by $\pi$ the Dynkin diagram automorphism of $A_1^{(1)}$-type 
defined by 
\begin{equation*}
\pi x_i = x_{i+1}\pi \quad (i=0,1, \; x=e,f,h), \quad 
\end{equation*}
where the subscripts are understood as elements of $\IZ/2\IZ$. 
We extend the affine Weyl group $W(A_1^{(1)})$ by adding the element
$\pi$ that satisfies algebraic relations 
\begin{equation}
 s_0^2 = s_1^2 = \pi^2 = 1, \quad \pi s_0 = s_1 \pi, \quad
\pi s_1 = s_0 \pi.
\label{extendedrel}
\end{equation}
We denote the extended Weyl group by $\widetilde{W}(A_1^{(1)})$. 
In the level-$0$ realization of the Chevalley 
generators \eqref{level0realization1},
the automorphism $\pi$ are realized by 
the adjoint action of the matrix
\begin{equation}
 \begin{pmatrix} 0 & z^{-1/2} \\ -z^{1/2} & 0 \end{pmatrix}.
\label{level0pi}
\end{equation}

As in the case of $s_j$, 
the action of $\pi$ on $g(0)$ induces a transformation on 
solutions of the $\pa$NLS equation through the Gauss decomposition, 
\begin{align*}
\lefteqn{\exp\! \left[ \sum_n t_nH_n \right] \pi^{-1} g(0) \pi
     = \pi^{-1} g(-t) \pi}\qquad\nonumber\\
 &= \left\{\pi^{-1} g^{[\mathrm{p}]}_{<0}(-t) \pi\right\}^{-1}
        \pi^{-1} g^{[\mathrm{p}]}_{\ge 0}(-t)\pi.
\end{align*}
It follows that 
\begin{equation*}
 \pi : \left\{\begin{array}{l}
         q(t)\mapsto -r(-t), \quad r(t)\mapsto -q(-t), \\
      \phi(t)\mapsto -\phi(-t), \quad a(t)\mapsto -b(-t), 
                                \quad b(t)\mapsto -a(-t).
\end{array}\right.
\end{equation*}
The sets of transformations 
$\langle s_0^\mathrm{L},s_1^\mathrm{L},\pi\rangle$ and 
$\langle s_0^\mathrm{R},s_1^\mathrm{R},\pi\rangle$
satisfy the relation of the extended affine Weyl 
group \eqref{extendedrel}
and thus we have obtained two different realizations of 
$\widetilde{W}(A_1^{(1)})$.

\section{Similarity reduction and monodromy problem}
In this section, we formulate a similarity condition 
of soliton equations in algebraic framework 
and consider the relation to monodromy problem of
a linear ordinary differential system.

\subsection{Similarity condition for soliton equation}
First, we impose a constraint for the 
initial data $g(0) = g(z;0)$: 
\begin{equation}
[d, g(z;0)] 
= \alpha H_0 g(z;0) + \beta g(z;0) H_0 + \gamma g(z;0)K.
\label{sim_g(0)}
\end{equation}
Here $\alpha, \beta, \gamma$ are complex parameters.
This relation leads to the following constraint 
for $g(t) = g(z;t)$ of \eqref{defg(t)}:
\begin{align}
[d, g(z;t)] &= 
\left(\alpha H_0 
+ \sum_{n>0} nt_n \frac{\partial}{\partial t_n} \right)g(z;t) 
+ \beta g(z;t) H_0 + \gamma g(z;t) K,
\label{sim_g(z;t)} 
\end{align}
because the generators of the homogeneous Heisenberg
subalgebra $H_n$ satisfy the condition, 
\begin{equation}
 \exp\left( \sum_{n>0} t_n H_n \right) 
\cdot d \cdot
 \exp\left( \sum_{n>0} t_n H_n \right) 
= d - \sum_{n>0} n t_n H_n.
\label{exptnHnd}
\end{equation}
Note that $d = z\pa_z$ is the 
derivation for the homogeneous gradation.
These conditions correspond to the similarity conditions
for $g(z;0)$ and $g(z;t)$:
\begin{align*}
 &g(\lambda z;0) 
= \lambda^{\alpha H_0} g(z;0) \lambda^{\beta H_0},\\
 &g(\lambda z;t) 
= \lambda^{\alpha H_0} g(z;\tilde{t}) \lambda^{\beta H_0}, 
\quad
\tilde{t} \defeq (\lambda t_1, \lambda^2 t_2, \ldots)
\end{align*}
by taking the exponential with respect to $\lambda$ of
the operators of both hand 
side of \eqref{sim_g(0)}, \eqref{sim_g(z;t)} respectively.

By applying the
Gauss decomposition to $g(z;t)$ with respect to the principal gradation, 
we obtain a constraint for $g^{[\mathrm{p}]}_{<0}(z;t)$ and
$g^{[\mathrm{p}]}_{\ge 0}(z;t)$ such as
\begin{align}
[d, g^{[\mathrm{p}]}_{<0}(z;t)] &= 
[\alpha H_0, g^{[\mathrm{p}]}_{<0}(z;t)]
+ \sum_{n>0} nt_n 
\frac{\partial g^{[\mathrm{p}]}_{<0}(z;t)}{\partial t_n},
\label{sim_g<0} \\
[d, g^{[\mathrm{p}]}_{\ge 0}(z;t)] &= 
\left(\alpha H_0 + \sum_{n>0} nt_n B_n \right)
g^{[\mathrm{p}]}_{\ge 0}(z;t) 
 + \beta g^{[\mathrm{p}]}_{\ge 0}(z;t) H_0 
 + \gamma g^{[\mathrm{p}]}_{\ge 0}(z;t) K.
\label{sim_g>=0}
\end{align}
These conditions correspond to
the similarity conditions:
\begin{equation}
 g^{[\mathrm{p}]}_{<0}(\lambda z; t) = 
  \lambda^{\alpha H_0} g^{[\mathrm{p}]}_{<0}(z;\tilde{t})
  \lambda^{-\alpha H_0}, 
\quad
 g^{[\mathrm{p}]}_{\ge 0}(\lambda z;t) =
  \lambda^{\alpha H_0} g^{[\mathrm{p}]}_{\ge 0}(z,\tilde{t}) 
\lambda^{\beta H_0}
\label{simgp}
\end{equation}
Especially, the first few components of $g^{[\mathrm{p}]}_{<0}(z;t)$
\eqref{formalExpansion}
and $g^{[\mathrm{p}]}_{\ge 0}(z;t)$ \eqref{expansion_=0} satisfy the
following conditions:
\begin{gather}
 q(\tilde{t}) = \lambda^{-2\alpha-1}q(t), \quad
 r(\tilde{t}) = \lambda^{2\alpha}r(t), \quad
 \phi(\tilde{t}) = \left(\log\lambda^{-(\alpha+\beta)} \right)\phi(t),
\label{similarityqr} \\
 a(\tilde{t}) = \lambda^{2\beta}a(t), \quad
 b(\tilde{t}) = \lambda^{-2\beta+1}b(t).
\label{similarityab}
\end{gather}
\begin{proposition}
If we set
\begin{equation}
 M = \alpha H_0 + \sum_{n>0} nt_n B_n,
\label{defofM}
\end{equation}
then $M$ and $B_n$ $(n=1,2,\ldots)$ satisfy
the zero-curvature equations:
\begin{equation}
 \left[ z\frac{d}{dz} - M, 
        \frac{\partial}{\partial t_n} - B_n
 \right] = 0.
\label{compati}
\end{equation}
\end{proposition}
\begin{proof}
By the definition \eqref{defofM} of $M$ and  
relations \eqref{SatoEq<}, \eqref{SatoEq>=},
\eqref{sim_g<0}, \eqref{sim_g>=0}, we can describe
\begin{align*}
 z\frac{d}{dz} -  M 
 &= g^{[\mathrm{p}]}_{<0} 
\left(z\frac{d}{dz} - \alpha H_0 
       - \sum_{n >0}n t_n H_n \right) 
(g^{[\mathrm{p}]}_{<0})^{-1} \\
 &= g^{[\mathrm{p}]}_{\ge 0}  
\left(z\frac{d}{dz} + \beta H_0 \right)
(g^{[\mathrm{p}]}_{\ge 0})^{-1}. 
\end{align*}
Therefore, by multiplying $(g^{[\mathrm{p}]}_{<0})^{-1}$ from the
left and $g^{[\mathrm{p}]}_{<0}$ from the right to the formula
$$
 \left[ z\frac{d}{dz} - \alpha H_0 
       - \sum_{n >0}n t_n H_n, 
        \frac{\partial}{\partial t_m} - H_m
 \right] = 0 \quad (m=1,2, \dots)
$$
or
$$
 \left[ z\frac{d}{dz} + \beta H_0, 
        \frac{\partial}{\partial t_m} - H_m
 \right] = 0 \quad (m=1,2, \dots),
$$
we have the equation \eqref{compati}.
\end{proof}

\subsection{Monodromy problem and Painlev\'e IV}
We now fix a positive integer $l >0$ and
restrict the operator for the time evolution to $\exp[\sum_{n=0}^l t_nH_n]$,
or we put $t_{l+1} = t_{l+2} = \cdots = 0$ in \eqref{exptnHnd}. 
Then $M$ of \eqref{defofM} becomes a element of affine Lie algebra. 
Under the realization \eqref{level0realization1},
we get a system of linear differential equations for a
$2 \times 2$ matrix $Y = Y(z;t_1,\dots,t_l)$:
\begin{equation}
 z\frac{\partial}{\partial z} Y = M Y, \qquad
 \frac{\partial}{\partial t_n} Y = B_n Y
 \quad (n = 1, \dots, l).
\label{linearsystem}
\end{equation}
This linear problem defines a monodromy preserving deformation
of linear ordinary differential system,
with regular singularity at $z=0$ and irregular singularity 
of rank $l$ at $z=\infty$.
We regard $t_1, t_2, \dots, t_l$ as a deformation parameter at $\infty$,
and $\alpha, \beta$ as monodromy data at $\infty$, $0$ respectively.

Hereafter, we set $l=2$ and put $t = t_2 = 1/2$.
Then $M$ of \eqref{defofM} for $B_1$ \eqref{B_1} and $B_2$ \eqref{B_2} 
can be written as
\begin{align}
 M =& H_2 + (-2qE_1 + 2rF_2) + (x + 2qr)H_1 \notag \\
    &\quad + (-(2xq + q')E_0 + (2xr -r')F_1) + (\alpha + k)H_0,
\label{matrixM}
\end{align}
where
\begin{equation}
 k = 2xqr + q'r-qr' - 2q^2r^2 
   = \left( x\frac{\partial}{\partial x} 
     + \frac{\partial}{\partial t} \right) \phi(t).
\label{constantk}
\end{equation}
The second equality is given by \eqref{diffEq_for_phi}.
We set
\begin{equation}
 \varphi = 2qr, \quad
 \psi_1 = -2x - \frac{q'}{q}, \quad
 \psi_0 = 2x - \frac{r'}{r}.
\label{fhensu}
\end{equation}
The compatibility condition \eqref{compati}
for the linear system \eqref{linearsystem}
for the restricted $M$ of \eqref{matrixM} and $B_1$ of \eqref{B_1}
present the
following system of differential equations:
\begin{align}
   \varphi' &=  -\varphi(\psi_1 + \psi_0),  
\label{y'} \\
    \psi_1'  &= \psi_1(2\varphi + \psi_1 + 2x) - 4\beta, 
\label{f1'}\\
    \psi_0'  &= \psi_0(-2\varphi + \psi_0 - 2x) -2(2\beta - 1).
\label{f0'}
\end{align}
In addition, the similarity condition for $\phi$ of \eqref{similarityqr} 
fix the value of $k$ \eqref{constantk}:
\begin{equation}
  k = -\alpha - \beta.
\label{valofk}
\end{equation}
So we have the relation
\begin{equation}
 \psi_0 - \psi_1 - \varphi + \frac{2(\alpha + \beta)}{\varphi} = 2x.
\label{souwa}
\end{equation}
\begin{proposition}
Each of the quantities $\varphi, \psi_1$ and $-\psi_0$ solve
the fourth Painlev\'e equation \eqref{Painleve4} with
the following parameters:
$$
\begin{array}{c|cc}
          &  \nu_1   &  \nu_2 \\ \hline 
 \varphi  & \alpha - 3\beta + 1 &  -2(\alpha + \beta)^2  \\
 \psi_1   & -2\alpha - 1 &  -8\beta^2  \\
 -\psi_0  & 2\alpha &  -2(2\beta -1)^2  
\end{array}
$$
\end{proposition}
\begin{proof}
Differentiating \eqref{y'}, 
we obtain
\begin{equation*}
 \varphi'' = -\varphi'(\psi_0+ \psi_1)  
              -\varphi\left( 2(\varphi + x)(\psi_1 - \psi_0) 
              + \psi_1^2 + \psi_0^2 -8\beta +2 \right).    
\end{equation*}
Then, by using the relations \eqref{y'} and \eqref{souwa},
we have $\mathrm{P_{IV}}$ \eqref{Painleve4} for $\varphi$.
For $\psi_1$, the relations \eqref{y'}, \eqref{f1'} and \eqref{souwa} give
the equations,
\begin{align}
   &\psi_1'  = 2\varphi \psi_1 + \psi_1^2 + 2x\psi_1 - 4\beta, 
\label{f1y1} \\
   &(\varphi \psi_1)' = \frac{(\varphi \psi_1)^2}{\psi_1} 
             - \varphi \psi_1 \left(\frac{4\beta}{\psi_1} 
                + \psi_1 \right)
             + 2(\alpha + \beta)\psi_1.
\label{f1y2}
\end{align}
Then, differentiating the first equation \eqref{f1y1} and
eliminating $\varphi \psi_1$ by using \eqref{f1y1} and \eqref{f1y2},  
we have the fourth Painlev\'e equation for $\psi_1$.
The other case $-\psi_0$ can be treated in the similar way.
\end{proof}

\textbf{Remark 1.}
Ablowitz et al. presented the fourth Painlev\'e equation
as a similarity reduction of $\partial$NLS \cite{ARS}.
In our notation, their results correspond to 
the equation for $\varphi$.
However, their result has only one parameter $\beta$, 
and corresponds to the special case $\alpha = -1/4$.
We give 
the fourth Painlev\'e equation
with full parameters.

\textbf{Remark 2.}
Our system of linear equations
\eqref{linearsystem} is not a
special case of the generalized 
Painlev\'e systems given by Noumi and Yamada \cite{NY:A, ICM}.
Their system is based on the 
similarity reduction of the  
principal hierarchy.

\bigskip

Jimbo and Miwa \cite{JM3} showed that $\mathrm{P_{IV}}$ is 
obtained as a similarity reduction of the NLS equation.
Their results correspond to the Gauss decomposition
of homogeneous-type in our setting. 
Since $g^{[\mathrm{h}]}_{<0}(z;t)$ \eqref{homog<0} and
$g^{[\mathrm{h}]}_{\ge 0}(z;t)$ \eqref{homog>=0}
satisfy the same similarity conditions as \eqref{simgp}:
\begin{equation*}
 g^{[\mathrm{h}]}_{<0}(\lambda z; t) = 
  \lambda^{\alpha H_0} g^{[\mathrm{h}]}_{<0}(z;\tilde{t})
  \lambda^{-\alpha H_0}, 
\quad
 g^{[\mathrm{h}]}_{\ge 0}(\lambda z;t) =
  \lambda^{\alpha H_0} g^{[\mathrm{h}]}_{\ge 0}(z,\tilde{t}) 
\lambda^{\beta H_0},
\end{equation*}
so the solutions 
of the NLS equation \eqref{NLS} satisfy the conditions
\begin{equation*}
 q(\tilde{t}) = \lambda^{-2\alpha-1}q(t), \quad
 \hat{r}(\tilde{t}) = \lambda^{2\alpha-1}\hat{r}(t).
\end{equation*}
By the same discussion as above, 
in the level-$0$ realization \eqref{level0realization1}
of $\tilde{B_1}$ \eqref{B_1tilde} and $\tilde{B_2}$ \eqref{B_2tilde},
we have the linear problem, 
\begin{equation*}
 \frac{\partial}{\partial z} Y = A(z) Y, \qquad
 \frac{\partial}{\partial x} Y = B(z) Y, 
\end{equation*}
with
\begin{align*}
 A(z) &= z \begin{pmatrix} 1 &  0 \\ 0 & -1 \end{pmatrix}
   + \begin{pmatrix}    x      & -2q \\ 
                    2\hat{r} &  -x \end{pmatrix}
  + z^{-1}\begin{pmatrix}
        \alpha + 2q\hat{r} & -2xq - q' \\
     2x \hat{r} - \hat{r}'  & -(\alpha + 2q\hat{r})
    \end{pmatrix},\\
 B(z) &= z \begin{pmatrix}
           1  &  0  \\
           0  & -1
       \end{pmatrix}
     + \begin{pmatrix}
             0      &  -2q  \\ 
         2\hat{r} &  0
    \end{pmatrix}.
\end{align*}
These can be identified with the result of \cite{JM3}.

Furthermore, we use 
\eqref{fhensu}, \eqref{souwa} and \eqref{rhat} to 
show that 
\begin{equation}
\label{relation_varphi_phi_1}
  \varphi \psi_1 = 4q\hat{r} + 2(\alpha + \beta). 
\end{equation}
Applying the relation \eqref{relation_varphi_phi_1} to 
the compatibility condition
$$
 \left[ \frac{\partial}{\partial z} - A(z), 
  \frac{\partial}{\partial x} - B(z) \right] = 0, 
$$
we have \eqref{f1y1}, \eqref{f1y2} and thus obtain
the fourth Painlev\'e equation for $\psi_1$. 

\subsection{Relations to Hamiltonian system}

In \cite{Okamoto}, Okamoto showed that
the fourth Painlev\'e equation \eqref{Painleve4} is equivalent 
to the Hamilton system, 
$$
 y' = \frac{\partial H}{\partial p}, \quad
 p' = -\frac{\partial H}{\partial y}, 
$$
with the polynomial Hamiltonian, 
$$
 H = yp^2 - y^2p - 2xpy -2\theta_0 p + 2\theta_\infty y.
$$
This is represented as the following system of equations for $y$ and $p$:
\begin{equation}
\begin{array}{l}
 y' = y(2p - y - 2x) - 2\theta_0,      \\
 p' = p(2q - p + 2x) - 2\theta_\infty.
\end{array}
\label{canonical}
\end{equation}
Our system \eqref{y'}--\eqref{f0'} can be identified with 
\eqref{canonical} in two different ways. 
Firstly, 
if we eliminate $\psi_0$ from \eqref{y'} by using \eqref{souwa}, 
we have 
\begin{align*}
  \varphi' &= \varphi(-2\psi_1 -\varphi -2x) + 2(\alpha + \beta), \\
   \psi_1'    &= \psi_1(2\varphi + \psi_1 + 2x) - 4\beta, 
\end{align*}
which are equivalent to \eqref{canonical} with
$$
  (y,p) = (\varphi, -\psi_1), \quad 
  (\theta_0, \theta_\infty) = (-\alpha -\beta, 2\beta). 
$$ 
Secondly, if we eliminate $\psi_1$, we have 
\eqref{canonical} with
$$
  (y,p) = (-\psi_0, -\varphi), \quad
  (\theta_0, \theta_\infty) = (\alpha + \beta, 2\beta -1).
$$

\subsection{Weyl group symmetry for the fourth Painlev\'e equation}
\label{SymP4}

To construct a Weyl group symmetry for
the similarity solution of the $\pa$NLS 
hierarchy, 
we examine a similarity conditions 
for $s_j^{-1}g(0)$ and $g(0)s_j$ $(j=0,1)$.
We have
\begin{align*}
&[d, s_i^{-1}g(0)] 
  =\left\{ \left(-\alpha-\frac{1}{2} \right)H_0 + \frac{h_i}{2} \right\}s_i^{-1}g(0) 
    + \beta s_i^{-1}g(0)H_0 + \gamma s_i^{-1}g(0)K, \\
&[d, g(0)s_i] 
  = \alpha H_0 g(0)s_i 
   +g(0)s_i \left\{ \left(-\beta + \frac{1}{2}\right)H_0 - \frac{h_i}{2} \right\}
   +\gamma g(0)K
\end{align*}
by using the relations \eqref{sim_g(0)}
and
\begin{equation*}
 [d, s_i] = \frac{1}{2}h_is_i - \frac{1}{4}[H_0,s_i]
\qquad (i=0,1).
\end{equation*}
Therfore, we have two types of Weyl group actions for the
parameters $\alpha, \beta$ 
\begin{equation*}
\left\{ \begin{aligned}
&s_0^\mathrm{L} : 
\alpha \mapsto -\alpha - 1,\quad
\beta  \mapsto \beta,      \quad
\gamma \mapsto \gamma + \frac{1}{2}, \\
&s_1^\mathrm{L} : 
\alpha \mapsto -\alpha, \quad
\beta  \mapsto \beta,   \quad
\gamma \mapsto \gamma,
\end{aligned}
\right.
\end{equation*}
\begin{equation*}
\left\{
\begin{aligned}
&s_0^\mathrm{R} : 
\alpha \mapsto \alpha,     \quad
\beta  \mapsto -\beta + 1, \quad
\gamma \mapsto \gamma - \frac{1}{2}, \\
&s_1^\mathrm{R} : 
\alpha \mapsto \alpha, \quad
\beta  \mapsto -\beta, \quad
\gamma \mapsto \gamma.
\end{aligned}
\right.
\end{equation*}

Next we consider the right-action of the affine Weyl group under
the similarity condition \eqref{sim_g(0)}.
Applying the relation \eqref{valofk} to 
\eqref{diffEq_for_f0} and \eqref{diffEq_for_f1}, 
we have
\begin{align*}
\left( x\frac{\pa}{\pa x} + \frac{\pa}{\pa t} \right)\tilde{\psi}_0
  &= 2xr - r' + 2(\alpha + \beta) \tilde{\psi}_0, \\
\left( x\frac{\pa}{\pa x} + \frac{\pa}{\pa t} \right) \tilde{\psi}_1
  &= -2xq - q' - 2(\alpha + \beta) \tilde{\psi}_1.
\end{align*}
On the other hand, left-hand-side of these equations can
be written in
$$
 \left( x\frac{\pa}{\pa x} + \frac{\pa}{\pa t} \right)\tilde{\psi}_0
= (2\alpha + 1)\tilde{\psi}_0, \qquad
 \left( x\frac{\pa}{\pa x} + \frac{\pa}{\pa t} \right)\tilde{\psi}_1
= -2\alpha \tilde{\psi}_1
$$
by using \eqref{ftilde},
\eqref{similarityqr} and
\eqref{similarityab}.
Then, under the similarity condition,
$\tilde{\psi}_0$ and $\tilde{\psi}_1$ can be expressed as
\begin{equation}
\tilde{\psi}_0 =  \frac{2xr-r'}{1-2\beta} 
= \frac{r\psi_0}{1-2\beta}, 
\qquad
\tilde{\psi}_1 = \frac{-2xq-q'}{2\beta} 
= \frac{q\psi_1}{2\beta}.
\label{relofpsi}
\end{equation}
We remark that in the realization \eqref{level0realization1},
the right-action of the
affine Weyl group is represented as a
compatibility of the 
gauge transformation 
$$
s_0 Y = \left(1 - \frac{1-2\beta}{r\psi_0}E_{-1}\right)Y,
\qquad 
s_1 Y = \left(1 - \frac{2\beta}{q\psi_1}F_0 \right)Y
$$
for the linear system \eqref{linearsystem}.
This transformation is the same as 
the Weyl group symmetry of Painlev\'e type equation
given by Noumi and Yamada \cite{ICM}.

In the level-$0$ realization \eqref{level0realization1},
the action of the extended affine Weyl group can be
obtained in the same manner.
The matrix \eqref{level0pi} satisfies the 
condition
$$
 [d, \pi] = -\frac{1}{4} [H_0, \pi], 
$$
and then the relation 
\begin{align*}
   [d, \pi^{-1} g(0) \pi] 
&= \left\{ \left(-\alpha - \frac{1}{2} \right)H_0 \right\} 
   \pi^{-1}g(0)\pi + \pi^{-1}g(0)\pi
   \left\{ \left(-\beta + \frac{1}{2} \right)H_0 \right\}
\end{align*}
holds.
Therefore the action of $\pi$ for the
parameters is given by
\begin{equation*}
\pi : 
\alpha \mapsto -\alpha - \frac{1}{2},\quad
\beta  \mapsto -\beta + \frac{1}{2},\quad
\gamma \mapsto \gamma.
\end{equation*}

\subsection{Schlesinger transformations and discrete Painlev\'e
equations}
\label{subsec:SchlesingerTr_dP}

In the level-$0$ realization \eqref{level0realization1},
we consider the local solutions of the linear 
system \eqref{linearsystem} at $z=\infty$ and $z=0$.
They are obtained by the following formal serieses:
\begin{align*}
Y^{(\infty)}(z;t) 
&= g^{[\mathrm{p}]}_{<0}(z;t) 
\exp \left((-\alpha \log z^{-1})H_0 + \sum_{n = 1}^l t_nH_n\right),\\ 
Y^{(0)}(z;t) 
&= g^{[\mathrm{p}]}_{\ge 0}(z;t) 
\exp\big((-\beta\log z) H_0 \big).
\end{align*}
Here $g^{[\mathrm{p}]}_{< 0}(z;t)$ and
$g^{[\mathrm{p}]}_{\ge 0}(z;t)$ are the
solutions of $\pa$NLS hierarchy with the
similarity conditions \eqref{sim_g<0}
\eqref{sim_g>=0} and 
the set of parameters $(-\alpha, -\beta)$ corresponds
to the monodromy exponents.
The Schlesinger transformation relates the
two solutions $Y$ and $Y'$ of the isomonodromy 
problem for the equation at hand corresponding to
different sets of parameters.
The change in parameters $(-\alpha, -\beta)$ are
integers or half-integers.

In the case of Painlev\'e IV, 
the Schlesinger transformation can be understood
in terms of the extended affine Weyl group.
If we consider the 
transformation $s^{\mathrm{R}}_1 \pi s^{\mathrm{L}}_1$,
the parameters $(-\alpha, -\beta)$ are
transformed as
\begin{equation}
\label{s1Rpis1L}
s^{\mathrm{R}}_1 \pi s^{\mathrm{L}}_1: \;
(-\alpha, -\beta) \mapsto
\left(-\alpha + \frac{1}{2}, -\beta + \frac{1}{2} \right)
\end{equation}
which corresponds to the Schlesinger transformation.
Applying the realization \eqref{realization-of-Weyl}, 
\eqref{level0pi} of the extended affine Weyl group,
we can describe this transformation
as the compatibility condition of \eqref{linearsystem} 
with
$$
  M = \begin{pmatrix}
        z^2 +  (x+\varphi)z -\beta  & -2qz + q\psi_1  \\ 
        2rz^2 + r\psi_0z  & -z^2 -(x+\varphi)z + \beta
    \end{pmatrix}
$$
and
$$
 \overline{Y} = R Y, 
\qquad
 R = \begin{pmatrix} 0 & 0 \\ -r & 1 \end{pmatrix}z^{1/2}
 + \begin{pmatrix} 0 & 1/r \\ 0 & r/\tilde{\psi}_0 \end{pmatrix}z^{-1/2}.
$$
Note that $\tilde{\psi_0}$ is defined in \eqref{ftilde}.
So the transformation of $M$ is given by
$$
 \overline{M} = R M R^{-1} + z\frac{\pa R}{\pa z} R^{-1}.
$$
Note that by the composition of left-action and $\pi$,
the sign of the variable $x$ does not change.
Using the relations \eqref{souwa} and \eqref{relofpsi},
we obtain the image of $(\varphi, \psi_1)$ in
terms of $(\varphi, \psi_1)$:
\begin{align*}
 &\overline{\varphi}
= -2x - \varphi + \psi_0 
   + \frac{2(1-\beta)}{\psi_0},    \\
 &\overline{\psi_1} = -\psi_0.  
\end{align*}
We remark that $\psi_0$ can be written 
in $(\varphi, \psi_1)$ by \eqref{souwa}.
Putting $\varphi = -2\chi_n$, $\psi_1 = -2\omega_n$
(and $\overline{\varphi} = -2\chi_{n+1}, 
\overline{\psi_1} = -2\omega_{n+1}$)
we find
\begin{align*}
 &\chi_n + \chi_{n-1} = x - \omega_n + \frac{\beta}{\omega_n}, \\
 &\omega_n + \omega_{n+1} = x - \chi_n + \frac{\alpha + \beta}{2\chi_n}.
\end{align*}
These equations are reduced to the 
discrete Painlev\'e I equation \eqref{dPI}
by putting $X_{2n} = \omega_n$, $X_{2n-1} = \chi_n$ for
$n \in \IN$.

Note that 
the Schlesinger transformation
to another direction 
represented by $s^{\mathrm{R}}_1 s^{\mathrm{L}}_1 \pi$, 
$$
 s^{\mathrm{R}}_1  s^{\mathrm{L}}_1 \pi: \;
 (-\alpha, -\beta) \mapsto 
\left(-\alpha - \frac{1}{2}, -\beta + \frac{1}{2} \right)
$$
also gives the discrete Painlev\'e I equation
for $\psi_1$ and $\varphi - 4\beta/\psi_1$.

\section{Tau-functions and special solutions}
In this section, we consider the basic representations
of $\slh_2$ \cite{FK} to introduce ``$\tau$-functions''. 
Let $|\varpi_j\rangle$ be a highest weight vector 
associated with the highest weight $\varpi_j$ ($j=0,1$), i.e., 
\begin{equation*}
\begin{array}{l}
e_i|\varpi_j\rangle=0, \quad 
h_i|\varpi_j\rangle=\delta_{ij}|\varpi_j\rangle \quad(i,j=0,1), \\
f_0|\varpi_1\rangle=f_1|\varpi_0\rangle=0. 
\end{array}
\end{equation*}
We denote by $L(\varpi_j)$ the basic representations 
with the highest weight $\varpi_j$, and by $L(\varpi_j)^*$ its dual space. 

First we construct a realization of $L(\varpi_0)\oplus L(\varpi_1)$ 
on the space 
\begin{equation*}
V=\IC[x_1,x_2,\ldots]\otimes
\left(\mathop{\oplus}_{n\in\IZ}\IC e^{n\alpha/2}\right), 
\end{equation*}
where $\alpha\in (\IC h_0\oplus\IC h_1)^*$ satisfies 
$\alpha (h_0) = -2$, $\alpha (h_1) = 2$. 
The representation $(\rho,V)$ is given as follows:
\begin{align*}
&\rho(H_j)\left(P(x)\otimes e^{n\alpha}\right)=
\begin{cases}
2\frac{\pa P(x)}{\pa x_j}\otimes e^{n\alpha} & (j\ge 1),\\
2nP(x)\otimes e^{n\alpha} & (j=0),\\
-jt_{-j}P(x)\otimes e^{n\alpha} & (j\le -1),
\end{cases}\\
&\rho(K)\left(P(x)\otimes e^{n\alpha}\right)
 = P(x)\otimes e^{n\alpha},\\
&\rho(d)\left(P(x)\otimes e^{n\alpha}\right)=
-\sum_{m=1}^{\infty}mx_m\frac{\pa P(x)}{\pa x_m}\otimes e^{n\alpha}. 
\end{align*}
To describe the action of $E_n$ and $F_n$, we introduce the 
generating series, 
\begin{equation*}
X(z) = \sum_{n\in\IZ}X_n z^{-n-1} \quad (X=E,F). 
\end{equation*}
Then the action of $E(z)$ and $F(z)$ is given by the following
operators (``vertex operators''): 
\begin{align*}
\rho(E(z)) &=
e^{\eta(x,z)}e^{-2\eta(\tilde{\pa}_x,z^{-1})}
\otimes e^{\alpha}z^{H_0}, \\
\rho(F(z)) &=
e^{-\eta(x,z)}e^{2\eta(\tilde{\pa}_x,z^{-1})}
\otimes e^{-\alpha}z^{-H_0}, 
\end{align*}
with 
\begin{equation*}
 \eta(x,z) \defeq \sum_{n=1}^\infty x_n z^n, \quad
 \eta(\tilde{\pa}_x,z^{-1}) \defeq \sum_{n=1}^\infty 
\frac{z^{-n}}{n}\frac{\pa}{\pa x_n}.
\end{equation*}

The representation $(\rho,V)$ is decomposed as follows:
\begin{equation*}
V = V_0 \oplus V_1, \quad 
V_j = \IC[x_1,x_2,\ldots]\otimes
\left(\mathop{\oplus}_{n\in\IZ}\IC e^{(n+j/2)\alpha}\right) \quad 
(j=0,1). 
\end{equation*}
It is shown that each of the representation $V_j$ ($j=0,1$) 
is isomorphic to $L(\varpi_j)$ \cite{FK}, 
where the highest weight vector is given by 
$|\varpi_j\rangle = 1\otimes e^{j\alpha/2}$. 

Next we prepare several results on symmetric polynomials \cite{Mac}. 
We denote by $p_j(z)$ the $j$-th elementary power-sum symmetric polynomial 
with respect to variables $z_1$,\ldots, $z_n$: 
\begin{equation*}
p_j(z) = z_1^j + z_2^j +\cdots + z_n^j. 
\end{equation*}
The Schur polynomial $S_\lambda(x)$, labeled by the partition 
$\lambda=(\lambda_1,\ldots,\lambda_n)$ is expressed by 
\begin{equation*}
S_\lambda (x)= \det\left[s_{\lambda_i-i+j}(x)\right]_{1\le i,j\le n}, 
\end{equation*}
where $s_n(x)$ is the $n$-th elementary Schur polynomial defined by 
\begin{equation*}
\exp[\eta(x,\lambda)] = \sum_{j=0}^\infty s_j(x)\lambda^j, 
\end{equation*}
and $s_n(x)=0$ if $n<0$.

We then introduce a scalar product in $\IC[x_1,\ldots,x_n]$: 
\begin{equation}
\label{ScalarProd0}
\langle P(x), Q(x)\rangle =\frac{1}{n}\,\mathrm{C.T.}\!
\left[P(x_j=\frac{p_j(z)}{n})Q(x_j=\frac{p_{-j}(z)}{n})
\Delta(z)\Delta(z^{-1})\right], 
\end{equation}
where $\mathrm{C.T.}[f(z)]$ denotes the constant term of 
$f(z)\in\IC[z_1^{\pm 1},\ldots,z_n^{\pm 1}]$ and 
$\Delta(z)=\prod_{1\le i<j\le n}(z_i-z_j)$, 
$\Delta(z^{-1})=\prod_{1\le i<j\le n}(z_i^{-1}-z_j^{-1})$. 
It is well-known that the Schur polynomials $\{S_\lambda\}$, associated with 
partitions $\{\lambda=(\lambda_1\ge\lambda_2\ge\cdots),
\lambda_j\in\IZ_{\ge 0}\}$, are pairwise orthogonal with respect to the 
scalar product \eqref{ScalarProd0}. 
The scalar product \eqref{ScalarProd0} induces a scalar product 
on $V=V_0\oplus V_1$: 
\begin{equation*}
\langle P(x)\otimes e^{a\alpha}, 
Q(x)\otimes e^{b\alpha}\rangle
= \delta_{ab}\langle P(x), Q(x)\rangle, 
\end{equation*}
where $P(x),Q(x)\in\IC[x]$ and $a,b\in\IZ/2$. 
Since the Schur polynomials forms an orthogonal basis of $\IC[x]$, 
an orthogonal basis of $V_0\oplus V_1$ is given by 
$\{S_\lambda(x)\otimes e^{a\alpha}\}_{\lambda,a}$. 

Following \cite{BtK1,tau}, we define $\tau$-functions associated with 
$g(t)$ of \eqref{defg(t)} as 
\begin{align*}
\tau_n^{(j)}(t) &= \langle 1\otimes e^{(n+j/2)\alpha}, 
g(t)(1\otimes e^{j\alpha/2})\rangle\\
&= \langle 1\otimes e^{(n+j/2)\alpha}, 
\{g_{<0}^{[\mathrm{p}]}(t)\}^{-1}
g_{0}^{[\mathrm{p}]}(t)
(1\otimes e^{j\alpha/2})\rangle. 
\nonumber
\end{align*}
We can express $q(t)$, $r(t)$ of \eqref{formalExpansion} 
in terms of the $\tau$-functions: 
\begin{equation*}
q(t) = -\frac{\tau_1^{(0)}(t)}{\tau_0^{(0)}(t)}, \quad
r(t) = -\frac{\tau_{-1}^{(1)}(t)}{\tau_0^{(1)}(t)}. 
\end{equation*}

As an example of concrete solutions, we construct polynomial-type 
$\tau$-functions, which are written in terms of the Schur polynomials. 
To this aim, we prepare the two lemmas: 
\begin{lemma}
\label{Lemma:WeylOrbit}
Let $n$ be an integer. We have 
\begin{equation*}
\rho(s_0s_1)^n (1\otimes e^{j\alpha/2}) = 
\epsilon_n (1\otimes e^{(n+j/2)\alpha}), \\
\end{equation*}
where $j=0,1$ and $\epsilon_n = 1$ or $-1$ 
depending on the value of $n$.
\begin{proof}
This is a direct consequence of Lemma 12.6 of \cite{Kacbook}. 
\end{proof}
\end{lemma}
\begin{lemma}
\label{Lemma:RectSchur} 
(cf. \cite{IY})
Let $k$ be a non-negative half-integer. 
We have the following expression of weight vectors: 
\begin{align*}
&\rho(e^{F_m})(1\otimes e^{k\alpha})
 = \sum_{n=0}^{2k-m}(-1)^{\frac{n(n+2m+1)}{2}}
S_{\square(n,2k-m-n)}(t)\otimes e^{(k-n)\alpha}, \\
&\rho(e^{E_m})(1\otimes e^{-k\alpha})
 = \sum_{n=0}^{2k-m}(-1)^{\frac{n(n-1)}{2}}
S_{\square(2k-m-n,n)}(t)\otimes e^{(n-k)\alpha}, 
\end{align*}
where the rectangular Young diagram $(k^n)$ is denoted by 
$\square(n,k)$. 
\end{lemma}
\begin{proof}
A proof can be given by same way as Theorem 1 of \cite{IY}. 
We omit the detail.
\end{proof}

Now we define $g_l(0)$ as 
\begin{equation}
\label{g_l(0)}
g_l(0) = 
\begin{cases}
e^{f_1}(s_0s_1)^l & (l\ge 0),\\
e^{f_0}(s_0s_1)^l & (l<0), 
\end{cases}
\end{equation}
for an integer $l$. 
Using Lemmas \ref{Lemma:WeylOrbit}, \ref{Lemma:RectSchur}, we can 
calculate the corresponding $\tau$-functions explicitly: 
\begin{align*}
\tau^{(0)}_n &= 
\begin{cases}
\epsilon_l (-1)^{\frac{(l-n)(l-n+1)}{2}}
S_{\square(l-n,l+n)}(t) & (l\ge 0),\\
\epsilon_l (-1)^{\frac{(n-l)(n-l-1)}{2}}
S_{\square(1-l-n,n-l)}(t) & (l< 0),
\end{cases}\\
\tau^{(1)}_n &= 
\begin{cases}
\epsilon_l (-1)^{\frac{(l-n)(l-n+1)}{2}}
S_{\square(l-n,l+n+1)}(t) & (l\ge 0),\\
\epsilon_l (-1)^{\frac{(n-l)(n-l-1)}{2}}
S_{\square(-l-n,n-l)}(t) & (l< 0).
\end{cases}
\end{align*}
These $\tau$-functions give rational solutions of the 
$\pa$NLS equation \eqref{DerNLS1}. 

Furthermore, straightforward calculations show that 
$g_l(0)$ of \eqref{g_l(0)} satisfies the reduction condition 
\eqref{sim_g(0)} with the following parameters: 
\begin{align*}
l\ge 0 &: \; \alpha = 0, \quad \beta = -l, \\
l<0 &: \; \alpha = -\frac{1}{2}, \quad \beta = \frac{1}{2}-l. 
\end{align*}
Hence we can perform the similarity reduction to the rational 
solutions given above and obtain rational solutions for the 
Painlev\'e IV.
In this case, the Schur polynomials $p_n(t)$ are degenerated to 
the Hermite polynomials $H_n(t)$: 
\begin{eqnarray*}
\lefteqn{\left.\exp(zt_1+z^2t_2+\cdots)
\right|_{t_1=x,\, t_2=1/2,\, t_3=t_4=\cdots=0}}
\\
&& = \exp(xz+z^2/2) = 
\sum_{n\in\IZ}H_n(t)z^n. 
\end{eqnarray*}

If we introduce discrete time evolution as \eqref{s1Rpis1L},
\begin{equation*}
g_l(0;n) = s_1^\mathrm{R}\pi s_1^\mathrm{L}(g_l(0))
= s_0^{-1} \pi^{-1} (g_l(0)) \pi s_1,
\end{equation*}
the corresponding rational solutions solve the discrete 
Painlev\'e equation \eqref{dPI} as discussed in 
the section \ref{subsec:SchlesingerTr_dP}. 
We remark that the rational solutions for the discrete Painlev\'e I 
\eqref{dPI} constructed in \cite{OKS} are essentially the same as the above.

\section{Concluding remarks}
We have formulated the hierarchy of the $\pa$NLS equation and 
introduced a systematic method for similarity reductions 
to Painlev\'e-type equations.
We used the fermionic representation of $\slh_2$ 
to construct rational solutions. 
We remark that the rational solutions can be expressed as 
ratio of Wronski-type determinants, which is discussed in \cite{KK}. 

As pointed out by Okamoto \cite{Okamoto}, the fourth Painlev\'e equation 
has the Weyl group symmetry of $\widetilde{W}(A_2^{(1)})$-type. 
Adler \cite{Ad} and Noumi and Yamada \cite{PIV} propose a new representation of
$\mathrm{P_{IV}}$, in which the $\widetilde{W}(A_2^{(1)})$ symmetries 
become clearly visible.
The Weyl group symmetry introduced in this article is 
isomorphic to $\widetilde{W}(A_1^{(1)})$, which dose not seems 
to be a subgroup of the $\widetilde{W}(A_2^{(1)})$-symmetry 
discussed in \cite{PIV,Okamoto}. 
To understand the relationship of our 
$\widetilde{W}(A_1^{(1)})$-symmetry to whole symmetry of 
$\mathrm{P}_\mathrm{IV}$, it seems that we need to consider 
a larger group that contain both $\widetilde{W}(A_1^{(1)})$ and 
$\widetilde{W}(A_2^{(1)})$ as individual subgroups. 

Though we limited ourselves to the $A_1^{(1)}$-case in this paper, 
our method may be applied to other type of affine Lie groups. 
For instance, in the case of $A_2^{(1)}$ non-standard hierarchy \cite{KIK},
we can obtain the fifth Painlev\'e equation with full parameters
as a similarity reduction of the modified Yajima-Oikawa equation.
We will dicuss this subject elsewhere. 

\section*{Acknowledgments}
The authors would like to thank Professors 
B.~Grammaticos, K.~Hasegawa, T.~Ikeda, 
G.~Kuroki, 
S.E.~Rao, K.~Takasaki, T.~Tsuchida, K.~Ueno, 
and R.~Willox for their interests and discussions. 
The first author is partially supported by 
by Rikkyo University Special Fund for Research, 
and by the Grant-in-Aid for Scientific Research (No.~14740123) from 
the Ministry of Education, Culture, Sports, Science and Technology.
The second author is partially supported by 
the 21st Century COE program of Tohoku University:
Exploring New Science by Bridging Particle-Matter Hierarchy. 


\end{document}